\documentclass[onecolumn,12pt]{article}
\usepackage{enumerate}
\usepackage{amssymb}

\newcommand {\ket}[1] {|{#1}\rangle}

\newcommand {\bra}[1] {\langle{#1}|}

\newcommand {\cl}{\mathcal}

\newcommand {\norm} [1] {\parallel #1 \parallel}

\newcommand {\beq} {\begin{equation}}
\newcommand {\eeq} {\end{equation}}
\newcommand {\mrm}{\mathrm}
\newcommand {\ovl} {\overline}

\begin{document}
\title{Universality and The Criterion $d$ in Quantum Key Generation}

\author{Horace P.~Yuen \\
Department of Electrical Engineering and Computer Science, \\
Department of Physics and
Astronomy, \\ Northwestern University, Evanston, IL, 60208, \\
Email: yuen@eecs.northwestern.edu}
\maketitle
\begin{abstract} The common security criterion $d$ in quantum key distribution is taken to solve the universal composability problem in quantum key distribution as well as providing good general quantitative security guarantee. In this paper it is shown that these are a result of an invalid interpretation of $d$. The general security significance of $d$ is analyzed in detail. The related issues of universality and attacker's side information are discussed.
\end{abstract}

PACS \#: 03.67Dd

\section{Introduction}
There have been considerable theoretical and experimental developments on the generation of a fresh (information-theoretically) secure key between two users via a protocol of the BB84 type \cite{bb84,gisin02}. The terminology of quantum key distribution (QKD) has been used more often than not and we will take it to be synonymous with quantum key generation, other terminology that has been employed include quantum key agreement and quantum key expansion. While there is another approach \cite{yuen09IEEE} to quantum key generation which does not involve information-disturbance tradeoff or intrusion level estimation, in this paper the term QKD will refer to BB84, while ``BB84'' is used in a wide sense that includes Ekert-type entanglement protocols as well as protocols that utilize other states than number states. QKD is interesting because if offers the possibility of fresh key generation with information-theoretic rather than complexity-based security, and thus would survive future developments in computational power including quantum computers.

A most important foundational problem in QKD is to develop necessary and sufficient conditions for the security of QKD protocols via relevant mathematical criteria. It is imperative that such a criterion must possess a clear empirical or operational meaning that bears directly on the intuitive but sufficiently precise notion of security a user would intend. In addition to mathematical correctness, it is also imperative that the proof of an unconditional security claim includes all possible attacks Eve may launch consistent with the model situation and the laws of physics and logic with all the side information Eve may possess taken into account. The security criterion and analysis must be scrutinized scrupulously, because, contrary to most other problems in physics and other empirical sciences, the security conclusion \emph{cannot} be established by an experiment or a simulation. While it is in principle possible but difficult to falsify a security conclusion by experiment or simulation, a direct analysis of the meaning of the criterion and whether all side information has been accounted for is a more logically clear and direct approach
for ascertaining the validity of a security claim.

In this paper we carry out such an analysis for the so-called universality security claim in QKD made via a criterion $d$ in \cite{renner05,renner08}. The ordinary security significance of this $d$ apart from universality will also be analyzed. It is important to scrutinize this criterion because it is currently the only one under which universality has supposedly been proven, and the interpretation given to it also makes it a very attractive security criterion in general. The nature of these claims can be briefly summarized as follows.

When the two users generate a key $K$ between themselves under the criterion $d \leq \epsilon$, according to \cite{renner05,renner08,rgk05,konig07} they would share a perfect key between them with probability $p$ that is at least $1-\epsilon$, i.e., with probability $p \geq 1- \epsilon$ the key $K$ can be considered identical to a uniformly distributed bit string $U$ which is independent of whatever Eve has in possession. This is clearly a very desirable state of affairs for ordinary security concerns if $\epsilon$ is sufficiently small. With probability at least $1-\epsilon$, it also clearly implies \emph{universality}, viz., it is secure in any arbitrary context. For instance, it guarantees that any bit of $K$ remains perfectly secret even if some other part of $K$ is given to Eve. Thus, the so-called \emph{composition problem} is solved from this universally composable security where $K$ remains secure in any subsequent use of it in different contexts. This and similar possible interpretations of $d$ will be analyzed, both for the universality issue and for its usual security significance. The related problem of Eve's side information, in particular her knowledge of the privacy amplification code and the error correcting code the users employ will be brought to bear on the security issues. It is concluded that $d$ does not possess universality and security significance as the above claim suggests, and not even in a weaker form. Thus, the universality problem is still open in QKD, as is also the problem of a proper security criterion apart from composability.

In Section 2 the general universality problem will be discussed, especially in relation to $d$. In Section 3, the incorrect meanings of $d$ will be analyzed, with specific counter-examples given on the invalid inferences for the significance of $d$. In Section 4, some correct meanings of $d$ are discussed and the problem of its quantitative security guarantee is elaborated for realistic protocols. The general side information issue is commented on in section 5 and some concluding remarks are given in Section 6.

\section{Universal Composability and the \\Criterion $d$}

In a direct one-way BB84 protocol, a user Alice sends another user Bob a sequence of random bits and they check a portion of it to assure that the error rate (QBER) is below a given threshold. Then an error correcting code (ECC) is employed on the rest, the sifted key, to obtain an error-free sequence between them with high probability. The resulting error-corrected key, to be called $K_c$, is passed through a hash function or privacy amplification code (PAC) to further reduce Eve's possible information with the final generated key $K$ as output. Eve attacks by setting her probe on the transmitted states before they reach Bob, waiting to get all the public exchange between Alice and Bob, and extracting her information on $K$ by measuring her probes.

The generated $n$-bit $K$ would have \emph{perfect security} if it is uniformly distributed on the key space of $2^n$ values and is independent of everything in Eve's  possession. The \emph{composition problem} arises that when $K$ is used in a specific context, e.g., in one-time pad encryption of a new data string, Eve may gain new information that may be combined with whatever in her possession to yield possibly some information on $K$ that she otherwise cannot obtain. The prime example is how \emph{partial key leakage} would affect the rest of the key. When $K$ is used as one-time pad, a known-plaintext attack would reveal some of the bits exactly. The composition problem is then whether these known bits of $K$ would increase Eve's information on the rest of $K$. The additional quantum problem is that Eve may hold her probe in quantum memory and choose the measurement with this side information and perhaps be then able to unlock much new information she could not otherwise obtain.

It is crucially important that we be precise in what we are after in this composition problem. Thus two important distinctions on such composition issue would be made. \textit{First}, it is clear that universality cannot obtain on arbitrary side information that may bear on $K$, just classically. The side information could be trivially what $K$ itself turns out to be, as a specific $k$. It could also be about how $k$ was generated from a classical source, which would reveal something about $k$. Even ``arbitrary context of use'' is too vague as not all such contexts can be characterized, or at least it is not clear how they may be characterized, by a single mathematical formulation, classically or quantum mechanically. Until the side information Eve may possess is explicitly specified, it is not possible to tell whether it may have something to do with the otherwise uniform key to her. An explicit illustration with the use of ECC and also PAC on $K_c$ will be given in Section 5.

Thus, it is more appropriate to separate the composition problem by the context in which $K$ is actually used. Generally, partial key leakage is always an important consideration because some information on part of $K$ when used as key can often be obtained by known-plaintext attacks also in other than one-time pad application. One needs to assign a \emph{specific} quantitative measure on the ``information about $K$'' that is leaked. Typically it would be some specific known bits in $K$ or some Shannon information on $K$ itself directly. This leads us to the next point.

\textit{Secondly}, there is a quantitative issue on how one wants to measure the information leak. Just in classical statistics that not all \emph{unknown parameters} can be modeled as random variables, not all side information $S$ is a \emph{random variable} to Eve. This may arise because $S$ may take on too many, in fact an infinite number of possible values with resulting infinite entropy, which for example would be the case if $S$ describes a complete procedure of how $K$ was physically generated. Assuming that $S$ is a random variable, Eve could then possibly obtain information on $K$ equal to $H(S)$ when one measures information by Shannon entropy $H(\cdot)$. However, if we fix attention on the partial key leakage problem, it is the rest of the $K$ that we are concerned with. Thus, we have the following situation. Consider the example of a two-bit $K=(k_1,k_2)$ in which Eve knows $k_1 \oplus k_2$. If $k_1$ is revealed to her she would know $k_2$. On the other hand, she has one-bit information on $K$ to begin with. How do we want to measure the effect of one-bit leak in this case? Eve has one bit to begin with and knows one bit afterward. Did she learn anything new? The case for a long $K$ is similar.

It seems it is meaningful to consider \emph{how much better Eve} could tell the rest of $K$ from the other leaked portion of it as a quantitative measure of composition security in this partial key leakage scenario, the comparison being made between the same portion of the key before and after another portion is leaked for a given quantitative measure. In the above two-bit example, $k_2$ by itself is completely random to Eve before she knows $k_1$ but it determines $k_2$ with probability one. The key is therefore not \emph{PKL-secure}, a terminology we choose to denote security under partial key leakage. Any quantitative measure of PKL-security can be introduced on the above basis with different operational significance, but whatever it is, a uniform key independent of Eve's possession must have full PKL-security.

In the quantum case there is the additional issue of lockable information \cite{konig07}, that a random variable side inforamtion $S$ may reveal to Eve more than $H(S)$ bits of information on $K$ which is impossible classically. From the above two-bit example it may be seen that PKL-security may be related to how much Eve knows about $K$ before the leakage. In \cite{benor05}, it was suggested that if Eve's optimal mutual information on $K$, called the accessible information $I_\mathrm{acc}$, is exponentially small in $n$ for large $n$, then the $n$-bit $K$ is composition secure according to their quantitative definition.  While the mathematical result in \cite{benor05} is correct, it was pointed out in \cite{konig07} via a counter-example with one-time pad use of $K$ that the result does not have the interpretation given in \cite{benor05} to guarantee their composition security.

Already in \cite{renner05,renner08}, a criterion $d$ was used to ``establish'' universality. This $d$ is actually also equivalent to one of several criteria discussed in \cite{benor05}. Specifically, let $\rho_E^k$ be the state in Eve's possession conditioned on a generated key value $k$, and let \beq \label{uniformstate}
\rho_U :=\frac {1}{|K|} \sum_k \ket{k}\bra{k}
\eeq
be the completely mixed uniform state on the $|K|=2^n$ orthonormal $\ket{k}$'s. It is assumed that the ``a priori'' probability of $K$ to Eve before she measures on her probe is uniform. Then \emph{the security criterion} $d$ is the trace distance
\beq \label{d}
d:= \frac{1}{2} \norm{\rho_{KE}-\rho_U \otimes \rho_E}_1,
\eeq
where
\beq
\rho_{KE}:= \frac {1}{|K|}\sum_k \ket{k}\bra{k} \otimes \rho_E^k
\eeq
and
\beq
\rho_E:= \frac {1}{|K|} \sum_k \rho_E^k.
\eeq

The trace distance $\norm{\rho - \sigma}_1$ between two states is related to the classical variational (statistical/Kolmogorov) distance $\delta(P,Q)$ between two probability distributions as follows \cite{renner05}. For any POVM or von Neumann measurement made on $\rho$ and $\sigma$ with resulting distribution $P$ and $Q$, $\norm{\rho -\sigma}_1 \leq \epsilon$ implies $\delta(P,Q) \leq \epsilon$ where
\beq \label{variational}
\delta(P,Q):= \frac{1}{2}\sum_{x \in \mathcal{X}}|P(x)-Q(x)|,
\eeq
and $P,Q$ are over the same range $\cl{X}$. Lemma 1 of \cite{renner05,konig05} states that for any distributions $P,Q,$ of random variables $X,X'$, there exists a joint distribution $P_{XX'}$ such that the marginal distributions are $P_X = P, P_{X'}=Q$, and
\beq
\mathrm{Pr}[X \neq X'] = \delta(P,Q).
\eeq
From this result it is concluded \cite{renner05}, [5, Prop 2.1.1] that when $d \leq \epsilon$, with probability $p \geq 1-\epsilon$ the real and the ideal situation of perfect security can be considered identical, where the ideal situation is one where $K$ is replaced by a uniformly distributed random variable $U$ which is independent of $\rho_E^k$. This statement is repeatedly made \cite{note1} and provides the following two very desirable consequences. Under $d \leq \epsilon$, with probability $p \geq 1 - \epsilon$ the key $K$ is universally composable (or at least so for partial key leakage) and it is the same as the uniform $U$ case to Eve for usual security apart from composition. Such a key is called $\epsilon$\emph{-secure}.

In the next section we will show that this conclusion is not valid and neither is another related weaker one. The significance of $d$ for both security with and apart from composition will be analyzed in detail in section 4.

\section{Analysis of Criterion $d$: What $d$ Does Not Guarantee}

Before we embark on a full analysis of the operational meaning and quantitative significance of the criterion $d$, let us note the meaning of the various quantum states involved. Eve's $k$-dependent probe states $\rho_E^k$ and their average $\rho_E$ have clear meanings. The key-determining state $\rho_U$ is in Bob's possession, and we can make this simplification used in \cite{konig07,konig05,scarani08} for the purpose of this paper instead of assigning first different keys $k_A$ and $k_B$ to Alice and Bob as in \cite{rgk05}. Then the joint state $\rho_{KE}$ of (3) on the state space $\cl{H}_B \otimes \cl{H}_E$ shared between Bob and Eve component-wise is an idealization that does not obtain in concrete realistic BB84 protocols. This is because many classical and macroscopic subsystems intervene between Bob and Eve and there is no single pure quantum state that governs both and known to anyone. Even if $\rho_E^k$ is re-interpreted to denote the state outside of Bob's key-determining register, such an overall pure state still does not obtain in a real protocol. On the other hand, if Bob \emph{only} measures on the basis $\{\ket{k}\}$ which is so intended in all the previous references, there is no harm and is perhaps mathematically convenient for some purpose to consider $d$ in the form of (2) involving the entangled state $\rho_{KE}$ of (3). In this situation, since Bob and Eve perform their ``local'' operations separately, the criterion $d$ is \emph{exactly} equivalent to
\beq \label{dequivalent}
d = \mrm{E}_k [\norm{\rho_E^k-\rho_E}_1],
\eeq
which is a condition on $\cl{H}_E$ alone. Equality of the right hand sides of (2) and  (\ref{dequivalent}) follows from lemma 2 of ref. \cite{renner05} directly, with $\mrm{E}_k$ the average over the $2^n$ possible values of $K$. The right-hand side of (7) is indeed one of the criteria proposed in \cite{benor05}. The entangled form (2) may give an illusion of being a more general criterion, but under the conditions just described that Bob measures only on the basis $\{\ket{k}\}$, it is actually not.

It is tempting, with the entangled form of $d$ in (2), to consider $\rho_{KE}$ as close to the product state  $\rho_U \otimes \rho_E$ when $d$ is small. If one interprets $d\leq \epsilon$ as meaning that $\rho_{KE}$ is ``basically'' $\rho_U \otimes\rho_E$, e.g., $\rho_{KE}$ is equal to $\rho_U \otimes \rho_E$ with probability at least $p \geq 1 - \epsilon$, then the criterion $d$ has the great significance discussed in Section 2 without any need for a justification via the variational distance through (5)-(6). It would guarantee a uniform key $U$ is obtained with probability $p \geq 1 -\epsilon$ which is universally composable with the same probability, a quite satisfactory security situation when $d$ is sufficiently small. Such an interpretation seems to be made in various
places \cite{note1} directly from the expression (2) independently of and in addition to the argument from subsequent variational distance obtained from a measurement via (6).

However, as in the case of classical probability distributions, a single number criterion (without coding a sequence into it) cannot so capture a whole distribution or a quantum state. This is brought out in \cite{yuen09IEEE,yuen06qcmc} and will be further discussed later. The small trace distance between two quantum states $\rho$ and $\sigma$ does \emph{not} imply $\rho$ is ``basically'' $\sigma$ -- it is simply a numerical measure that is useful for various purposes similar to $\delta(P,Q)$, but it does not guarantee certain empirical meaning one may want to attribute to the relation between $\rho$ and $\sigma$. Similar to the case of Eve's mutual information on the key, the $d \leq \epsilon$ or $\delta_E := \delta(P,U) \leq \epsilon$ criterion is fine when $\epsilon$ is \emph{sufficiently} small. The \emph{serious} problem  is that for long bit sequences, they have to be extremely small, as described in section 4. Thus, unless the previous interpretation \cite{note1} holds the criterion does not provide adequate quantitative security guarantee similar to the case of Eve's mutual information per bit \cite{yuen09IEEE}. Unfortunately, the interpretation does \emph{not} hold.

In the following  we will analyze the significance of $d$ and show that it is does \emph{not} have either of the following three consequences:
\begin{enumerate} [(i)]
\item Through (5)-(6), the generated key $K$ is equal to a perfect key $U$ with probability $p \geq 1 -d$ and is completely independent of Eve's knowledge and possessions.
\item With probability at least $1-d$, $\rho_{KE}$ is equal to $\rho_U \otimes \rho_E$ shared by Bob and Eve.
\item Any probability distribution $P$ Eve may obtain by a measurement on the probe has $\delta(P,U) \leq d$ for the uniform distribution $U$ of the same support as $P$.
\end{enumerate}
In Section 4 the possible security significance of $d$ in both the before-usage and composition contexts will be explored.

The conclusion (i) above is explicitly asserted and ``proved'' as described in section 2 of this paper, first in reference \cite{renner05} and later repeated in many other papers \cite{note1}. The proof relies on the \emph{existence} of a joint distribution that yields the marginal distributions $P$ and $Q$ and gives (6). However, to the extent it makes sense to talk about such a joint distribution, (i) would follow only if ``there exists'' is replaced by ``for every''. This is because since there is \emph{no} knowledge on such joint distribution, one cannot assume the most favorable case via ``there exists'' for security guarantee or general ``interpretation'' that ref. \cite{renner05,renner08,rgk05,konig07,konig05,scarani08} give.

Indeed, it is not clear at all what realistic meaning can be given or claimed for the realization of such a joint distribution, other than the independent case $P_{XX'} = P\cdot Q$. In such case, even if both $P$ and $Q$ are the same uniform distribution so that $\delta(P,Q)=0$, we have $\textrm{Pr} [X \neq X'] = 1 - \frac{1}{N}, N = 2^n$, and the two sides of (6) are almost as far apart as they can be since both are between $0$ and $1$. This provides a counter-example to the interpretation and security guarantee.
\\\\
\emph{Counter-Example to (i)}:\\

Let $X$ and $X'$ be independent in (6). Then (6) is violated (almost maximally).
\\\\
Moving onto (ii), one may observe that $\rho_{KE}$ cannot be ``basically'' the same as $\rho_U \otimes \rho_E$ in general if $\rho_U$ and $\rho_E$ are ``close'' to pure states. In particular, an actual product pure state $\rho_U \otimes \rho_E$ is not ``basically'' the same as an entangled state $\rho_{KE}$ of (3). The specific statement (ii) has the following consequence:-
\beq
\rho_{KE}=(1-d) \rho_U \otimes \rho_E + d \sigma_{KE}
\eeq
for some state $\sigma_{KE}$ and  we have used a fixed $d$ for simplicity instead of carrying along  $d \leq \epsilon$. The same (near)-pure states $\rho_U$ and $\rho_E$ falsifies (8) in general. A specific numerical example would show (8) to be invalid as follows.

Consider the binary case of one-bit $k$ where conditioned on $k \in \{0,1\}$, Eve has a pure probe state $\ket{k_0}$ or $\ket{k_1}$. It is readily computed that in this case $d=\frac{1}{4}\norm{\ket{k_0}\bra{k_0}-\ket{k_1}\bra{k_1}}_1$, and is thus bounded between $0$ and $1/2$. If Eve makes the optimum binary quantum detection on her probe, her probability of success is from \cite{helstrom76}
\beq \label{psuccess}
P_c = \frac{1}{2} + d.
\eeq Under the hypothesis (8), we have from total probability decomposition,
\beq
P_c \leq (1-d) \frac{1}{2} + d \cdot 1 = \frac{1}{2} + \frac{d}{2}.
\eeq
An exact computation from (8) with
\beq
\sigma_{KE}= 2^{-n} \sum_k \ket{k}\bra{k} \otimes \sigma_E^k
\eeq
shows that
\beq
P_c = \frac{1}{2} + \frac{d}{4} \norm{\sigma_E^{k_0} - \sigma_E^{k_1}}_1
\eeq
or $P_c \leq 1/2 + d/2$ consistent with (10). Thus, the more secure scenario (8) with $P_c$ bounded by (10) contradicts the actual $P_c$ of (9) obtainable in this example.

The following counter-example specifically pertains to composition also. Consider a two-bit $K$ with
\begin{eqnarray}
\rho_E^{00} = \sigma \otimes \rho_1, \hspace{1cm}\rho_E^{01}=\sigma\otimes\rho_2 \\
\rho_E^{10} = \sigma \otimes \rho_2, \hspace{1cm}\rho_E^{11}=\sigma\otimes\rho_1,
\end{eqnarray}
for a fixed $\sigma$ on the first qubit and general states $\rho_1,\rho_2$ on the second. It is readily computed that for (13)-(14),
\beq
d = \frac{1}{4}\norm{\rho_1 - \rho_2}_1
\eeq
analogous to the above single-bit example. According to (ii), the whole $2$-bit sequence has a probability $1-d$ of being the uniform $U$, which implies that with the same probability, knowledge of the first bit leak implies nothing about the second, which cannot be determined by Eve with a success probability $P_c > 1/2 +d/2$ as in (10). On the other hand, it is evident from (13)-(14) that knowledge of the first bit would imply the second bit can be determined with success probability $P_c = 1/2 +d$ as in (9). Thus, in this example there is no quantitative PKL-security according to (ii), or from (i) which gives  the same quantitative guarantee as (ii). Combining these two examples we have \\\\
\emph{Counter-Example to (ii):}\\\\
Let $\rho_E^k$ be given by (13)-(14). Then the PKL-security for composition given by (i) or (ii) is violated. Also, the single-bit security of the second bit by itself violates (ii) as in (9). \\\\

Finally, for (iii), we note that the condition $\delta_E := \delta(P,U)=d$ has the significance that for $K$ with distribution $P$, any $m$-bit subsequence $1 \leq m \leq n$, has a probability $p_m$ from $P$ that satisfies \cite{nielsenchuang}
\beq
|p_m - 2^{-m}| \leq d.
\eeq
Thus, a sufficiently small $\delta_E$ would have the desirable property of guaranteeing any deviation of $K$-subsequence probability from uniform to be small. However, $d \leq \epsilon$ does not imply (iii) on either a joint distribution from a measurement on $\cl{H}_B \otimes \cl{H}_E$ or on Eve's distribution from her measurement on $\cl{H}_E$. This is because $\rho_E$ is not generally the uniform state $\rho_U$ even when the range of $\{\rho_E^k\}$ has dimension exactly equal to $|K|=2^n$. In concrete protocols, the dimension of $\rho_E^k$ would exceed $|K|$ considerably, through key sifting, error correction and privacy amplification. It does \emph{not} seem possible to have any state $\rho_E$ that would yield the uniform distribution to E upon her general or even just projection-valued measurements.

Counter-examples can be constructed easily on (iii) from $\rho_E \neq \rho_U$. We can also have the same example (13)-(14) on (ii) on the distribution from measuring on $\cl{H}_E$ when $\sigma = \ket{a}\bra{a}$ and Eve measures $\{\ket{a}, \ket{b}\}, \bra{a}b\rangle=0$ on the first probe qubit and the eigenvectors of $\rho_1 - \rho_2$ on the second. It is readily computed that in this case
\begin{eqnarray}
\delta_E = \frac{1}{2}+2d,\hspace{0.5cm}  d &\geq& \frac{1}{4} \\
 = 1, \hspace{1.5cm} d &\leq& \frac{1}{4} .
\end{eqnarray}
Thus, $\delta_E \leq d$ is not satisfied.\\\\
\emph{Counter-Example to (iii):}\\\\
The above (17)-(18) or any $\rho_E \neq \rho_U$.\\\\

Note that $\delta(P,Q) \leq \epsilon$ does not imply that the distribution $P$ is equal to $Q$ with probability $p \geq 1 -\epsilon$, similar to the case of $\norm{\rho-\sigma}_1 \leq \epsilon$. That is, the following is \emph{not} a consequence of $\delta(P,Q)=\epsilon$ from (5)
\beq
P(x) = (1-\epsilon) Q(x) + \epsilon P'(x),
\eeq
where $P'$ is a distribution on $\mathcal{X}$ \cite{note2}. Similar to the quantum case (8), under (19) one may \cite{note3} infer that the distribution $Q$ is obtained with probability $1-\epsilon$. The difference between the two is then:
\begin{enumerate} [(a)]
\item Under (19), the random variable $X$ has the distribution $Q$ with probability at least $1-\epsilon$.
\item Under $\delta(P,Q) \leq \epsilon, |p_m - q_m| \leq \epsilon,$ where $p_m$ and $q_m$ are any $m$-subsequence probabilities of the random vector $X$ under $P$ and $Q$.
\end{enumerate}

For (a), with probability $p \geq 1 -\epsilon$  all subsequences of $X$ have exactly the probabilities given by $q_m$. Under (b),  each subsequence of $X$ has probability  $q_m \pm \epsilon$. Not only is there a quantitative difference which we have discussed in this section, there is a uniformity property for (a) that is not shared by (b). The previous interpretation \cite{note1} is similar to (a) while only (b) holds. As we have shown, such interpretation cannot be maintained. Indeed, it appears counter-intuitive that with a high probability Eve knows \emph{exactly} nothing about the generated key, that not even a tiny amount of information could be derived from her probe and public announcement.

\section{Security Significance of $d$}

The exact security significance of $d$ with and without composition is analyzed in the following. First we note the analog in $\delta$ of the equivalence between (2) and (7) for uniformly distributed $K$,
\beq
\mathrm{E}_k [\delta(P_E^k, P_E)]= \overline{d},
\eeq
where $\overline{d}$ is now given by
\beq
\overline{d} := \delta(P_{kk'}, U_k Q_{k'}).
\eeq
In (21), $P_{kk'}$ is the joint distribution of Bob's measured key $k$ and Eve's measurement result $k'$ on her probe $\rho_E^k$, $U_k$ the uniform distribution on $k$ and $Q_{k'}$ Eve's distribution from measuring on $\rho_E$. The equality of (20) follows from $P_{kk'}=P_{k'|k}U_k$ in (21) with $P_E^k (k')=P_{k'|k}$ and $P_E = \frac{1}{|K|} \sum_k Q_k$. Equation (21) implies, from (16),
\beq
|P_{kk'}- U_k Q_{k'}| \leq \overline{d}.
\eeq

There is no composition significance to $\overline{d} \leq \epsilon$, from $d \leq \epsilon$, similar to that of (i) or (ii) or (a) in section 3. Intuitively, $\overline{d} \leq \epsilon$ says from (22) that any joint distribution $P_{kk'}$ is, up to $\overline{d}$, given by the product distribution $U_kQ_{k'}$ for which Eve's measurement would yield a $k'$ that is independent of the actual $k$. Thus, her estimate of $k$ has to come from her possible a priori information on $k$ and whatever is allowed through $\overline{d}$ of (21). However, the strongest way this restriction comes about is in spelling out the definition of $\delta(P,Q)$ in (20)-(21) and not from (22). Thus, (20) shows that only on average over $K$ is $P_E^k$ carrying no information on $k$ up to $\overline{d}$, the individual (22) would yield a useless $|K|\overline{d}$ constraint on $P_{k'|k}$ for typically large $|K|$. The averaging allows that for an individual $k$ value the security could be much worse.

To explore further the significance of $d \leq \epsilon$ or $\overline{d} \leq \epsilon$, we strengthen (7) to
\beq
d_k := \norm{\rho_E^k - \rho_E}_1 \leq \epsilon, \hspace{1cm}\forall k.
\eeq
From the triangle inequality, it follows immediately from (23) that
\beq
\norm{\rho_E^{k_1} - \rho_E^{k_2}}_1 \leq 2\epsilon,\hspace{1cm} \forall k_1, k_2.
\eeq

For the composition problem, especially for PKL-security, knowledge of a subsequence of $K$ would restrict the possible $k$'s to a smaller set. Let $\overline{k}$ be known and $k'$the remaining $K$. Then for each possible $k'$, the state $\rho_E^{\overline{k}k'}$ still satisfies, from (23), $\norm{\rho_E^{\ovl{k}k'}- \rho_E}_1 \leq \epsilon$. From (24) the trace distance between any two such possible $k$'s remains bounded by the same $2\epsilon$. It may then be inferred that partial key leakage has not affected the rest of $K$ in so far as the security conditions (23)-(24) are concerned. However, as discussed in section 2 it is $k'$, the rest of $K$, that matters in such context and $k'$ may be more readily identifiable as follows.

Eve's probability distribution from any measurement is now reduced from $2^n$ to $2^{n-m}$ possibilities when $m$ bits are leaked. This would generally reduce her error probabilities even under (23) or (24), although it is not guaranteed to be so on a per bit basis when normalized relative to the $n-m$ bit uniform distribution. In any event, a further \emph{specific quantitative measure} that has direct \emph{operational} meaning, such as Eve's success probability in identifying $k'$, needs to be proposed and a general proof provided on how strong the protocol is against partial key leakage. The previous conclusion of universal composability is so convenient due to the invalid conclusion that the generated key is uniform and independent of Eve's probe with probability $1-d$.

The condition (23) or (24) also implies that Eve's discrimination of $k$ would be difficult for sufficiently small $\epsilon$ as all the $\rho_E^k$ are clustered together closely. Good error probability estimates for the multiple quantum state discrimination problem are available \cite{tyson09}, but they do not cover the case where the success probability is small. It appears interesting to relate (23) or (24) to Eve's optimal success probability of determining $K$, or parts of it, which would provide clear security significance to $d_k \leq \epsilon$. Such work could be attempted both for before-usage security and for partial key leakage.

The criterion $d \leq \epsilon$ is used in the security proof  as follows. An $\epsilon$-secure key $K$ of appropriate rate $n/n_0$ is guaranteed from the measured QBER for various protocols \cite{renner05,renner08,rgk05,scarani08} via bounds involving various $\epsilon$-smooth entropies. In \cite{scarani08}, numerical plots on the original BB84 protocol are given for $\epsilon=10^{-5} \sim 2^{-16}$ and $n_0 > 10^4$ with $n/n_0 > 0.1$. These results do not rule out very insecure possibilities as follows.

There is a serious quantitative problem of $\delta_E$ guarantee that applies to even $d_k \leq \epsilon$ and assuming reduction of $\rho_E$ to $U$ can be achieved. It was pointed out \cite{yuen09IEEE,yuen06qcmc} that under the constraint $I_E/n \leq 2^{-l}$ for Eve's mutual information per bit $I_E/n$ on $K$, there are distributions on $K$ Eve may possibly obtain that gives her maximum probability $p_1$ of identifying the \emph{whole} key $K$ as $p_1 \sim 2^{-l}$. The same distribution gives
\beq
\delta_E = 2^{-l}-2^{-n},
\eeq
which is $\sim 2^{-l}$ for $l$ up to a good fraction of $n$. The subsequence of $K$ may be obtained with higher probability under the $I_E/n$ constraint but it satisfies the more secure (16) under the $\delta_E$ constraint. Nevertheless, for $l \sim 10^2$ or smaller, a $n=10^3$ key is far from ``perfect'' not only for the whole $K$ but also for many subsequences of $K$ . In practice, only $l \sim 10$ for $I_E/n$ has been achieved experimentally \cite{hasegawa07} and it is not clear it can be made much better for $\delta_E$. The problem is that unless $l \sim m$, the guarantee from (16) or (25) that the $m$-subsequence probability $p_m \leq 2^{-l}$ is \emph{far worse} than that of a uniform key $K=U$. Even if $l \sim 20$ can be experimentally achieved, it is quite poor for $m \sim 100$ and $n \geq 100$ as compared to a uniform key.

In this connection, it may be pointed out that the incorrect meaning assigned to $d$ \cite{note1} makes the security situation appear much more favorable. Indeed, it seems $d=2^{-l}$ would play the same role as the message authentication key of $l$ bits used in creating the public channel, that except for a probability $p=2^{-l}$ the cryptosystem is secure. However, as a message authentication key of $l \gtrsim 50$ bits may be needed for say, $ n \gtrsim 10^3$ bits, a small $l \sim 20$ is not really a good guarantee. It is important to \emph{note} that many numbers involved in a concrete realistic protocol with $n \geq 100$ are exponentially small. It is necessary to compare two very small numbers carefully.

Note also that the average constraint $d \leq \epsilon$ is much weaker than $d_k \leq \epsilon$ of (23). From Markov's inequality \cite{cover91},
\beq
\textrm{Pr}[X \geq \delta] \leq \mathrm{E}[X]/\delta
\eeq
for a nonnegative valued random variable $X$, one may guarantee $X \geq \delta$ with a probability $\epsilon$ by imposing $\mrm{E}[X] \leq \epsilon \delta$ instead of just $\mrm{E}[X] \leq \epsilon$. For concrete protocol guarantee, a single application of (26) already severely strains the numerical requirement for a given quantitative level. The situation becomes much worse for multiple guarantees of the form (23) or (24).

In sum, we have shown  that $d$ does \emph{not} have the meaning attributed to it in the literature, while for concrete protocols its quantitative security significance with and without composition does not seem to be much better than the mutual information criterion.

\section{Side Information}

As discussed in Section 2, the composition problem involves side information Eve may obtain depending on the exact context in which the key $K$ is used. There are similar side information even during protocol execution which Eve obtains but is \emph{not} accounted for in protocol security analysis. This is partly due to the fact that the exact message authentication code for creating a tamper-proof public channel, the error correcting code, as well as the privacy amplification code are not usually precisely specified in security analysis, and when they are, their specific character are not usually taken into account. We will indicate some of the issues in this section which are related to our analysis in this paper. We will not discuss the message authentication problem here, except noting that the full specific details of the protocol could be exploited by an attacker as the following discussion on ECC and PAC demonstrates.

In the criterion $d$ of (2)-(4) the assumption is made that the a priori distribution of the key $K$ is uniform. However, the often suggested direct use of an ECC on the key $K_C$ to be corrected after sifting and testing would bias the a priori probability. This is because an ECC would ``decode'' $K_c$ to a message, but only perfect codes \cite{macwilliams77} would have equal size decision regions that are needed to assure equal a priori probabilities for the different $K$ values. However, perfect codes are rare and none is a good candidate for a concrete protocol, while the relative sizes of the decision regions with any decoding rule for common ECC's have not been studied in coding theory. This problem does not arise when the $K_c$ is used as additive noise to a uniformly chosen codeword.

Similarly, a matrix is typically chosen as a PAC, say an $m \times n$ Toeplitz matrix that belongs to a ``universal class'' \cite{renner05,renner08}. However, a singular matrix would leak information to Eve \emph{even} if the original bit sequence to be compressed is perfect. Indeed, in the binary case a rank $m-r$ matrix would leak $r$ bits of Shannon information to Eve as a linear combination of the $n$ bits that gives a $0$ would leak one bit of information. This problem has not been dealt with in the literature.

In particular, eq.~(11) in ref \cite{renner05} cannot guarantee an $\epsilon$-secure key for any given universal hash function. At best it could only do that when averaged over such functions. Even then it is rather amazing to have automatically such guarantee on a universal family of Toeplitz matrices that has many singular members. In any event, another application of Markov Inequality for individual guarantee is necessary in this case if one does not analyze further the proportion of singular matrices in a Toeplitz family.

The point here is that unless the full protocol is specified, one would not be able to tell how Eve may utilize any side information to obtain further information on the generated key. In particular, universality is too vague to allow a complete mathematical characterization and each specific context such as partial key leakage should be analyzed individually.

\section{Concluding Remarks}
We have shown that the criterion $d$, the seemingly most potent criterion that has so far appeared in the QKD literature, does not solve the composability problem such as partial key leakage. Its empirical and quantitative significance has been analyzed in detail, and found to be different from what it is so far taken to be. Not only does it not guarantee the generated key $K$ is uniformly distributed and independent of the attacker's probe with a high probability, it also does not guarantee the variational distance $\delta_E = \delta(P,U)$ between the attacker's possible distribution $P$ on $K$ and the uniform distribution $U$ is small. Furthermore, it appears very difficult for $d$ to provide adequate quantitative security guarantee through its legitimate meaning for concrete realistic protocols, similar to the situation of the attacker's mutual information per bit criterion \cite{yuen09IEEE}. The general issue of side information that Eve may obtain is also discussed in relation to both universality and general quantitative security. It appears there are much to be elaborated on QKD security analysis in regard to both the empirical significance of security criteria and their quantitative adequacy in concrete realistic protocols.

\section{Acknowledgement}
I would like to thank R.~Nair, M.~Raginsky, and R.~Renner for useful discussions. This work was supported by AFOSR.


\begin{thebibliography}{10}
\expandafter\ifx\csname url\endcsname\relax
  \def\url#1{\texttt{#1}}\fi
\expandafter\ifx\csname
urlprefix\endcsname\relax\def\urlprefix{URL }\fi

\bibitem{bb84} C.H. Bennett and G. Brassard, in {\it Proc. IEEE
                Int. Conf. on Computers, Systems, and Signal
                Processing}, Bangalore, India (IEEE, Los Alamitos,
                CA), 175-179 (1984).

\bibitem{gisin02} A general review can be found in
               N. Gisin, G. Ribordy, W. Tittel, H. Zbinden,
               Rev. Mod. Phys. {\bf 74}, 145-195 (2002).

\bibitem {yuen09IEEE}H.P.~Yuen, to appear in IEEE J. Sel. Top. in Quantum
Electronics, also arXiv: http://arxiv.org/abs/0906.5241

\bibitem{renner05} R.~Renner, and R.~Konig, Second Theory of Cryptography Conference (TCC), Lecture Notes in Computer Science, vol. 3378 (Springer, New York, 2005), pp. 407-425.

\bibitem{renner08} R.~Renner, Int. J. Quant. Inf. 6, 1 (2008); also arxiv.org: quant-ph/0512258.

\bibitem{rgk05} R.~Renner, N.~Gisin, and B.~Kraus, Phys. Rev. A 72, 012332 (2005).

\bibitem{konig07} R.~Konig, R.~Renner, A.~Bariska, and U.~Maurer, Phys. Rev. Lett. 98, 140502 (2007).

\bibitem{benor05} M.~Ben-Or, M.~Horodecki, D. W.~Leung, D.~Mayers, and J.~Oppenheim, Second Theory of Cryptography Conference (TCC), Lecture Notes in Computer Science Vol. 3378 (Springer, New York, 2005), pp. 386–406.

\bibitem{konig05} R.~Konig, U.~Maurer, and R.~Renner, IEEE Trans. Inform. Theory 51 (2005), p. 2381-2401.

\bibitem{note1} This is explicitly stated in [4] -- p.~14, [5] -- Section 2.2.2, [6] --p.~ 012332-5, [7]--p.~140502-3, [11], p.~200501 -2.

\bibitem{scarani08} V.~Scarani and R.~Renner, Phys. Rev. Lett. 100, 200501 (2008).

\bibitem{yuen06qcmc} H.P.~Yuen in: O.~Hirota, J.H.~Shapiro, M.~Sasaki (Eds.), Proceedings of the QCMC, NICT Press, 2006, p.~163.

\bibitem{helstrom76} C.W. Helstrom, {\it Quantum Detection and
                Estimation Theory}, Academic Press, New York (1976).

\bibitem{nielsenchuang} M.A.~Nielsen and I.L.~Chuang, \emph{Quantum Computation and Quantum Information}, Cambridge University Press, 2000; p.~401.

\bibitem{note2} A specific counter-example was provided to the author by R.~Renner.

\bibitem{note3} There is the subtle issue of ensemble identity similar to the ''partition ensemble fallacy'' in quantum mechanics that may cast doubt on such inference on distributions, which we do not enter into here.

\bibitem{tyson09} J.~Tyson, J. Math. Phys. 50, 062102 (2009).

\bibitem{cover91} T.M.~Cover and J.A.~Thomas, \emph{Elements of Information Theory}, Wiley, 1991.

\bibitem{hasegawa07} J.~Hasegawa, M.~Hayashi, T.~Hiroshima, A.~Tomita,  Asian Conference on Quantum Information Science 2007,  Shiran-kaikan, Kyoto, Sep.3-6, (2007).

\bibitem{macwilliams77} F.J.~MacWilliams and N.J.A.~Sloane, \emph{The Theory of Error Correcting Codes}, New York, Elsevier/North Holland, 1977.
\end{thebibliography}
\end{document}